\documentstyle[12pt]{article}

\def\hybrid{\topmargin -20pt    \oddsidemargin 0pt
        \headheight 0pt \headsep 0pt
        \textwidth 6.25in       % A4 paper
        \textheight 9.5in       % A4 paper
        \marginparwidth .875in
        \parskip 5pt plus 1pt   \jot = 1.5ex}

\hybrid

\catcode`\@=11

\def\marginnote#1{}

\newcommand{\r}[1]{(\ref{#1})}

\begin{document}

\null\vskip-24pt \hfill KL-TH 01/02 \vskip-10pt \hfill \texttt{hep-th/0103082%
} \vskip0.3truecm

\begin{center}
\vskip 3truecm {\Large \textbf{Holographic Trace Anomaly and Cocycle of Weyl Group}}\\[0pt]
\vskip 1truecm %\vfill
{\large \textbf{R. Manvelyan,}}$^{*\#}${\large \textbf{\ \footnote{\textbf{%
email:\texttt{manvel@physik.uni-kl.de} }}}} {\large \textbf{ R. Mkrtchyan, }}$^{*}$%
{\large \textbf{\ \footnote{\textbf{email:\texttt{mrl@arm.r.am}}}
}}\\[0pt]{\large \textbf{and H. J. W. M\"uller-Kirsten}}$^{\#}$
{\large \textbf{\ \footnote{\textbf{%
email:\texttt{mueller1@physik.uni-kl.de} }}}}
\vskip 1truecm %\addresses
$^{*}$\textit{Department of Theoretical Physics,\\[0pt]
Yerevan Physics Institute \\[0pt]
375036 Yerevan, Armenia}

\vskip 0.5truecm %\addresses
$^{\#}$\textit{Department of Physics, Theoretical Physics\\[0pt]
University of Kaiserslautern, Postfach 3049 \\[0pt]
67653 Kaiserslautern, Germany}\\[0pt]
\end{center}

\vskip 2truecm
\begin{abstract}
The behavior of the divergent part of the bulk AdS/CFT
effective action is considered with
respect to the special finite diffeomorphism transformations
acting on the boundary as a Weyl transformation of the boundary
metric. The resulting 1-cocycle of the Weyl group is in full
agreement with the 1-cocycle of the Weyl group obtained from the
cohomological consideration of the effective action of the
corresponding CFT.
\end{abstract}
\newpage

\section{Introduction}
The conformal anomaly plays an important role in the investigation of
non-renormalization properties of supersymmetric theories in
different dimensions. Investigation of the structure of the conformal or
trace anomaly has a 20 years history \cite{Duff}, but its general
structure was first discovered in \cite{DSCH} using a
scaling property of the effective action.The cohomological properties
of trace anomalies were considered in \cite{Bon, KMM} and the connection
with 1-cocycles of the Weyl group was established. This cohomological
object is important because it can be considered as the full effective
action in a conformal flat background. The structure of the conformal
anomaly can be considered in a general way using the Wess-Zumino
consistency condition \cite{WZ} and cocyclic properties of
the effective action, the nonlocal terms of which generate the anomaly
after local Weyl variation. The exact values of coefficients of
different terms in the anomaly could be fixed by calculation in
perturbation theory using the free field representation.

The AdS/CFT correspondence \cite{Malda} provides an important tool
for the calculation of conformal anomaly coefficients and
conformal correlation functions in the large N limit of ${\cal
N}=4$ Super-Yang-Mills in $d=4$ \cite{arkady} and superconformal
theories in other dimensions. In this approach the anomaly appears
after classical calculation of the on-shell $d+1$ dimensional
effective action in the form of logarithmically divergent terms
\cite{HS}. Thus the AdS/CFT approach offers the possibility to
investigate the effective action in the strong coupling limit. The
Weyl transformation properties and some cohomological
considerations in this approach were presented in \cite{Theis}.
The authors of the latter discovered the relation between the
infinitesimal Weyl transformation of the boundary metric and a
certain one-parameter family of $d+1$ dimensional diffeomorphisms.
The finite form of these diffeomorphisms was obtained in \cite{S}.
In this note we reproduce the 1-cocycle of the Weyl group in
$d=2,4$  from investigation of the behavior of the divergent part
of the AdS/CFT effective action with respect to these finite
diffeomorphism transformations. We show that a cut-off
regularization of the pure bulk contribution in the effective
action (without boundary term) is responsible for the anomaly and
the cocycle generation after applying the diffemorphism
transformation, and correctly reproduces the 1-cocycle of the Weyl
group (analogue of the Liouville action) in $d=2$ and $4$
dimensions. In Section 2 we review the cohomological structure of
the effective action and the derivation of 1-cocycles of the Weyl
group. In Section $3$ we derive this cocycle from the
Heningson-Skenderis approach using finite diffeomorphism
transformations of bulk divergent terms.

\section{Cocycle of the Weyl Group}
In this section we review the cohomological  properties of the CFT
effective action in the external gravitational field. This
consideration is based on the results of \cite{Rieg} , \cite{KMM}.

We consider the effective action for $d$-dimensional
conformal matter $\varphi$
 in an external gravitational field:
\begin{equation}
\label{1} W(g) = \ln \int D_{g}\varphi \exp\{ -S_{CFT}(\varphi;
g)\}
\end{equation}
where $ S_{CFT}(\varphi;g)$ is the classical Weyl and diffeomorphism
invariant action for matter fields and where the Weyl transformation
is defined as:
\begin{equation}
\label{2} g_{ij} \rightarrow e^{2\sigma(x)} g_{ij} ,
 \qquad  \varphi \rightarrow e^{\Delta\sigma(x)}\varphi .
\end{equation}
Here $\Delta$ is the conformal weight of the matter field. Then, for an
infinitesimal $\sigma$, we can write the equation for the anomaly
\begin{equation}\label{3}
\delta_{\sigma} W(g) = \int T^{i}_{i} \delta\sigma(x) \sqrt{g}
d^{2k}x .
\end{equation}
The Wess-Zumino consistency condition in the case of Weyl
transformations is simply a statement of the symmetry of the second
conformal variation of the effective action:
\begin{equation}\label{4}
\frac{\delta ^{2}W(g)}{\delta\sigma(x)\sqrt{g}\delta\sigma(y)} =
\frac{\delta ^{2}W(g)}{\delta\sigma(y)\sqrt{g}\delta\sigma(x)}
\end{equation}
 or, in other words,
\begin{equation}\label{5}
\frac{\delta {\cal A}(x)}{\delta\sigma(y)} =\frac{\delta {\cal
A}(y)}{\delta\sigma(x)}
\end{equation}
where we set $T^{i}_{i}={\cal A}(x)$.

The considerations of \cite{DSCH, KMM} lead to the following
general structure of the solution of the Wess-Zumino consistency condition
in all (even) dimensions. For any local function of the metric $A(g)$
(i.e. the anomaly) the WZ consistency condition provides  the
following statement, concerning the structure of $A(g)$:

1) ${\cal A}(g)$ is a sum  of the
following terms with arbitrary coefficients:

a)Type {\bf A}- Euler density,

b)Type {\bf B}- Weyl-invariant polynomials over the Riemann tensor and
its covariant derivatives,

c) Covariant total derivatives of polynomials over the Riemann tensor
and its covariant derivatives.

We can add the following comment:

2) The latter type of the anomalies are the Weyl variations of local
functionals of the metric. Taking into account the fact, that the definition
of the measure in the functional integral can always be changed by
multiplying the measure by an exponential of the local
functionals (counterterms) of the metric, one can deduce, that the third
(i.e. c)) type of solutions of the WZ condition are in that sense inessential
and might be called trivial below.

So we can classify the possible anomalies in d=2 and d=4 in the
following way:

1) In $d=2$ there is only a type {\bf A} anomaly,
\begin{equation}\label{6}
{\cal A}_2=-\frac{c}{24\pi}R
\end{equation}
where $c$ is the central charge of the corresponding $CFT_2$;

 2) In d=4 there are nontrivial type {\bf A} and {\bf B}
anomalies,
\begin{equation}\label{7}
{\cal A}_4= \alpha E_4 +\beta I_4
\end{equation}
where $E_{4}$ is the Euler density and $I_4$ the square of the Weyl
tensor,
\begin{eqnarray}\label{8}
  E_4 =R^{ijkl}R_{ijkl}-4R^{ij}R_{ij}+R^2 ,\\
  I_4 =-C^{ijkl}C_{ijkl}=-R^{ijkl}R_{ijkl}+2R^{ij}R_{ij}-\frac{1}{3}R^2
.\label{9} \end{eqnarray}
The constants $\alpha$ and $\beta $ depend upon the
specific mode content of the theory and interaction.

We now consider the change of the measure in the
functional integral for the conformal matter field $\varphi$ in the
external gravitational field under the finite Weyl transformation
\r{2}.

 The measure in the functional integral changes in the following way:
\begin{equation}\label{10}
  D_{e^{2\sigma(x)}g}\varphi = D_{g} \varphi \exp{S(\sigma;g)}
\end{equation}
This type of relation is very important since it is, for example,
the starting point for the calculation of the critical exponent of
$2d$ gravity \cite{DDK}.

 The action $S(\sigma; g)$ in \r{10} has to satisfy some conditions.
First, in the case of infinitesimal transformations
$\delta\sigma(x)$ it has to reproduce the trace anomaly:
\begin{equation}\label{11}
S(\delta\sigma(x); g_{ij}) = \int T^{i}_{i} \delta\sigma(x)
\sqrt{g} d^{2k}x
\end{equation}
Second,  $S(\sigma; g)$  has to satisfy the following property,
which follows from the application of \r{10} to the composition of
two Weyl transformations $\sigma_{1}$ and $\sigma_{2}$:
\begin{equation}\label{12}
S(\sigma_{1} + \sigma_{2}; g) - S(\sigma_{1}; e^{2\sigma_{2}}g) -
S(\sigma_{2}; g)= 0
\end{equation}
which means that  $S(\sigma; g)$  is the 1-cocycle of the group of
Weyl transformations \cite{FS}.

On the other hand, the action  $S(\sigma; g)$ coincides with the
finite variation of the anomalous effective action, due to the
properties  \r{1} and \r{10}. In other words
\begin{equation}\label{13}
S(\sigma, g) =  W(e^{2\sigma} g) - W(g) ,
\end{equation}
and non-triviality of the cocycle $S(\sigma;g)$  follows from the
fact that $W(g)$ is a non-local, $Diff(2k)$-invariant  functional of
$g_{\alpha\beta}$ in the case of type {\bf A} and {\bf B}
anomalies.
 Thus, we can easily calculate the trivial cocycles (type c)) as a coboundary of
 local counterterms   $W_{0}(g)$ which we shall call from now on
 0-cochains:
\begin{equation}\label{14}
S_{0}(\sigma,g) = \triangle W_{0}(g) ,
\end{equation}
where we have defined the coboundary operator $\triangle$ on
0-cochains as the finite Weyl variation \r{13}. Then we can define
1-cochains as local functions $W_{1}(\sigma,g)$ of group
parameter and metric with coboundary operator:
\begin{equation}\label{15}
\triangle W_{1}(\sigma_{1},\sigma_{2},g) = W_{1}(\sigma_{1} +
\sigma_{2}; g)
 - W_{1}(\sigma_{1}; e^{2\sigma_{2}}g) -  W_{1}(\sigma_{2}; g)
\end{equation}
It is easy to see that $\triangle^2 = 0$ which is exactly the
cocyclic property \r{12}. One can generalize this construction on
higher cohomologies of the Weyl group \cite{FS}.

The nontrivial cocycles can be obtained from  the solution of eq.
\r{12} with condition \r{11}. To obtain the solution we have to take
$\sigma_2 =\sigma$ and $\sigma_1 =\delta\sigma$ and get the differential form
of \r{12}:
\begin{equation}\label{16}
 \delta S(\sigma; g) = S(\delta\sigma; e^{2\sigma}g)
= \int A(R(e^{2\sigma}g))\delta\sigma \sqrt{g} d^{2k}x .
\end{equation}
The explicit form of the solution for the two-dimensional case is the
well-known Liouville  action \cite{Pol}
\begin{equation}\label{17}
S_{d=2}(\sigma,g)=\frac{c}{24\pi}\int d^2
x\sqrt{g}(g^{ij}\partial_i \sigma
\partial_j \sigma - R \sigma) .
\end{equation}
 We can restore this cocycle in the usual way using the variation of
the non-local effective action:
\begin{equation}\label{18}
S_{d=2}(\sigma,g)=\triangle
\frac{-c}{96\pi}\int\sqrt{g}R\frac{1}{\sqrt{g}\Box}\sqrt{g}R .
\end{equation}
In four dimensions the explicit form of the cocycle, corresponding to
$E_4$, was first found in \cite{Rieg},
\begin{eqnarray}\label{19}
S_{E}(\sigma,g)&=&\int d^4 x\sqrt{g}\left( 2(\nabla_i \sigma
\nabla^i \sigma)^2 +4\nabla_i \sigma \nabla^i \sigma \nabla^2
\sigma\right.\nonumber\\ &
&\left.-4(R^{ij}-\frac{1}{2}g^{ij}R)\nabla_i \sigma \nabla_j
\sigma - \sigma E_4\right) .
\end{eqnarray}
For type {\bf B} anomaly in $d=4$ there is a linear nontrivial
cocycle corresponding to the single invariant density
$C_{ijkl}C^{ijkl}$:
\begin{equation}\label{20}
S_{C}(\sigma,g) = \int C_{ijkl}C^{ijkl}\sigma (x) \sqrt{g}d^4x .
\end{equation}
This expression satisfies the cocyclic property \r{12} and can
appear in the Weyl transformation of the measure \r{10}. All other
freedom in the definition of the cocycle of the Weyl group is connected with
the special choice of local counterterms in the regularized effective
action. These counterterms will generate trivial cocycles
corresponding to the type c) full derivative contribution in
the anomaly. In d=4 we have only one local counterterm with
independent Weyl variation and therefore here  we can obtain only
one trivial cocycle:
\begin{equation}\label{21}
S_{0}(\sigma,g) = \triangle\int R^2\sqrt{g}d^4x .
\end{equation}
As was shown in \cite{Rieg} this trivial cocycle can be used to
reduce the nontrivial one connected with the Euler density up to the second
order in $\sigma$ with the fourth-order conformal-invariant
differential operator acting on a scalar field of zero conformal
weight as a kinetic term and then can be expressed as a Weyl
variation of some nonlocal effective action (analog of Liouville
theory in d=4). The same but much more complicated structure exists
also in d=6 \cite{KMM} and d=8 \cite{Ans}. Finally we want to
describe the cocycle of the Weyl group for ${\cal N}=4$ Super-Yang-Mills
theory. It is well known that the anomaly of this theory vanishes for
a Ricci-flat background \cite{HS} and has the following form:
\begin{equation}\label{22}
{\cal A} = \alpha \left(E_{(4)} + I_{(4)}
\right)=2\alpha\left(\frac{1}{3}R^2 - R^{ij} R_{ij}\right) ,\quad
\alpha = -\frac{N^2-1}{64\pi^2} .
\end{equation}

We can easily solve our equation \r{16} and obtain the cocycle for the
d=4, ${\cal N}=4$ SYM anomaly or we can consider this cocycle as a sum
of \r{19} and \r{20} corresponding to Euler density and
Weyl invariant $I_4$,
\begin{eqnarray}\label{23}
S_{\alpha (E+I)}(\sigma,g)&=&2\alpha\int d^4 x\sqrt{g}\left[(\nabla_i \sigma
\nabla^i \sigma)^2
+2\nabla_i \sigma \nabla^i \sigma \nabla^2 \sigma\right.\nonumber\\
& &\left.-2(R^{ij}-\frac{1}{2}g^{ij}R)\nabla_i \sigma \nabla_j
\sigma -\left(\frac{1}{3}R^2 - R^{ij} R_{ij}\right)\sigma\right] .
\end{eqnarray}

\section{Holographic Effective Action and Cocycle}

We now try to derive the latter cocycle from the Heningson-Skenderis
AdS  construction. It is well known that the effective action of all
maximally supersymmetric theories in d=3,4, and 6 can be derived
from the on-shell d+1 dimensional  gravitational action with
asymptotically AdS classical solution \cite{HS}:
\begin{eqnarray}
W &=&\frac 1{2k_{d+1}^2}\int_{M_{d+1}}d^{d}xd\rho \sqrt{G}\left( {\bf R}+%
2\Lambda \right) -\frac
1{2k_{d+1}^2}\int_{\partial M_d}d^d\sigma \sqrt{\gamma }2K ,\label{24} \\
\Lambda &=&-\frac{d(d-1)}2 ,\qquad K=D_{\mu}n^\mu  . \label{25}
\end{eqnarray}

The boundary term (with $\gamma$ the induced metric, $K$ the trace
of the extrinsic curvature and $n^\mu$ a unit vector normal to the
boundary) is necessary in order to obtain an action which depends
only on first derivatives of the metric and to obtain a
well-defined variational problem with Dirichlet boundary
conditions for the usual Einstein equations. The metric $G_{\mu
\nu }$ is degenerate on the boundary and the boundary metric is
defined up to conformal transformations. To obtain the anomalous
effective action following standard procedures \cite{HS} we have
to solve the equation of motion for this action using the
Fefferman-Graham \cite{FG} coordinate system for $d+1$ dimensional
metrics $G_{\mu\nu}$:
\begin{eqnarray}
ds^2 &=&G^{\mu \nu }dx^\mu dx^\nu =\frac{d\rho ^2}{4\rho ^2}+\frac{%
g_{ij}(\rho ,x)}\rho dx^idx^j , \label{26} \\
\mu ,\nu &=&\rho ,1,2,... d;\qquad i,j=1,2,...d ,\nonumber\\
g(x,\rho)&=&g_{(0)} + \cdots + \rho^{d/2} g_{(d)} + h_{(d)}
\rho^{d/2} \ln \rho + \cdots .\label{27}
\end{eqnarray}
The equations of motion determine $g_{(n)}(x)$ in terms of
$g_{(0)}(x)$:
\begin{eqnarray}\label{28}
  &&g_{(2)ij}={1 \over d-2}(R^{(0)}_{ij} - {1 \over 2(d-1)}
  R^{(0)}g_{(0)ij}) ,\\
  &&g_{(4)ij}= \cdots .\nonumber
\end{eqnarray}
We then have to insert this expansion of $g$ into the on-shell
gravitational action \r{24} and perform the integration over
$\rho$. However, the on-shell action diverges because the boundary
metric is degenerate. We therefore  we have to regularize the
action using a restriction on the $\rho$ integral with some
infrared cut-off $\rho \geq \varepsilon $, and evaluate the
boundary term at $\rho =\varepsilon $. Then
\begin{equation}\label{29}
S = \frac 1{2k_{d+1}^2}\int d^{d}x\int_{\rho \geq
\varepsilon}d\rho\frac{d}{\rho^{{d\over 2} +1}} \sqrt{g(x,\rho )}
-\frac{1}{k_{d+1}^2}\int_{\partial M_d^\varepsilon}d^d\sigma
\sqrt{\gamma }K .
\end{equation}
The resulting on-shell effective action contains divergences as
poles $1\over \varepsilon^n$, a logarithmic divergence and a
finite part (which is the essential effective action if we apply
the minimal renormalization scheme):
\begin{equation}\label{30}
  W_{reg} = {1 \over 2k^{d+1}} \int d^d x \sqrt{\det g_{(0)}} \left(
\varepsilon^{-d/2}{\bf a}_{(0)} + \ldots  - \ln \varepsilon\, {\bf
a}_{(d)} \right) + W_{finite} .
\end{equation}
From this expression we can easily recognize the anomalous
behavior of the unknown  $W_{finite}$ by investigating the behavior of
the divergent terms with coefficients ${\bf
a}_{(n)},\,\,n=0,1,....,d$ with respect to scale transformation of
the metric \cite{HS}, $\delta g_{(0)}=2\delta\sigma g_{(0)}$ and
$\delta \varepsilon =2\delta\sigma \varepsilon$, with {\it
constant} $\delta\sigma$. Taking into account that the entire
action \r{30} is invariant and all negative powers of $\varepsilon
$ terms are scale invariant, one can derive from variation of the
logarithmically divergent term that the holographic anomaly is:
\begin{equation}\label{31}
{\cal A} = -{1 \over k^{d+1}}{\bf a}_{(d)} .
\end{equation}
The explicit expressions of the logarithmically divergent terms
\cite{HS, KSS} in $d=2$ and $d=4$ are: ${\bf a}_{(2)}=\frac{1}{2}
R$ and ${\bf a}_{(4)}=-R^{ij} R_{ij}/8 + R^2/24$. The value of
the gravitational constant in $d=5$ can be obtained from \cite{kleb},
\begin{equation}\label{32}
  \frac{1}{2\kappa_5^2}=\frac{N^2}{8\pi^2} ,
\end{equation}
where $N$ is the number of coincident $D3$ branes. It is easy to
see that inserting this expression into \r{32}, we can recognize
the anomalous coefficient of ${\cal N}=4$ SYM in the large N
limit. The origin of this anomaly generation is also well-known
\cite{Theis}. There are special $d+1$ dimensional diffeomorphisms
of asymptotic AdS space-time parameterized with one scalar
function $\sigma(x)$ acting on the boundary as a Weyl
transformation of the metric $g_{(0)}$.These diffeomorphisms leave
the form of the Fefferman-Graham metric invariant. Solving this
condition
 we can write down the following important
formulas for such diffeomorphisms \cite{S} in the form of power
series in $\rho'$:
\begin{eqnarray}
\rho&=&\rho' e^{-2 \sigma(x')} + \sum_{k=2} a_{(k)}(x') \rho'^k,
\qquad x^i= x'^i + \sum_{k=1} a_{(k)}^i(x') \rho'^k ,\label{33}\\
a_{(2)}&=&-\frac{1}{2}(\partial \sigma)^2 e^{-4 \sigma}, \qquad
a_{(3)}={1 \over 4} e^{-6 \sigma} \left( {3 \over 4} (\partial
\sigma)^2 +
\partial^i \sigma \partial^j
\sigma g_{(2)ij} \right), \label{34}\\
a^i_{(1)}&=&\frac{1}{2} \partial^i \sigma e^{-2 \sigma}, \,
a^i_{(2)}=-\frac{1}{4} e^{-4 \sigma} \left(\partial_k \sigma
g_{(2)}^{ik} + \frac{1}{2} \partial^i \sigma (\partial \sigma)^2 +
\frac{1}{2} \Gamma_{kl}^i \partial^k \sigma \partial^l \sigma
\right) .\label{35}
\end{eqnarray}
The nice property of these diffeomorphisms is that the
transformation rules for $g_{(n)}$ are the usual Weyl
transformations of metric $g_{(0)}$:
\begin{eqnarray}
g_{(0)ij}' &=& e^{2 \sigma} g_{(0)ij} ,\label{36}\\
 g_{(2)ij}' &=& g_{(2)ij} +\nabla_i \nabla_j \sigma -
 \nabla_i \sigma \nabla_j \sigma + \frac{1}{2} (\nabla\sigma)^2
 g_{(0)ij} ,\label{37}\\
 g_{(2)ij}'&=& \cdots .\nonumber
\end{eqnarray}
The important point here is that the anomalies are described only
by the logarithmic divergence originating only from the bulk
integral. The interesting point about the boundary extrinsic
curvature term is the following: {\it there is no contribution
from this term in the anomaly, because there is no $\rho $
integral there, and there is no $\varepsilon $ independent $R^2$
order contribution from this term }. We can easily check this from
the definition of the boundary term at the point
$\rho=\varepsilon$ \cite{HS}:
\begin{equation}\label{38}
{1 \over 2k^{d+1}}\int_{\partial M_d}d^d\sigma
\sqrt{\gamma}2D_{\mu}n^\mu ={1 \over 2k^{d+1}}\int d^dx{1\over
\rho^{d/2}} \left(-2 d \sqrt{\det g(x,\rho )} +4 \rho\partial_\rho
\sqrt{\det g (x,\rho )}\right)|_{\rho=\epsilon} ,
\end{equation}
using the trivial identity
\begin{equation}\label{39}
  (-2d +4 \rho\partial_\rho)\rho^{d/2}=0 .
\end{equation}
In addition we can note that the coefficient $h_{(d)}$ in the
expansion of the metric is traceless and there is no room for a
contribution in the anomaly from this term also. The absence of an
$\varepsilon$ independent term is important because this $R^2$
type term could be considered like a local counterpart to the
effective action and produce after Weyl variation a trivial ($\Box
R$) contribution in the anomaly. We can now try to apply the
special diffeomorphisms \r{33} to the divergence part of the
effective action \r{30}, and we can expect to obtain our cocycle
\r{23} constructed earlier from the solution of the cohomological
equation \r{12}. Before doing that we note that the definitions of
boundary and boundary term are invariant with respect to $d+1$
dimensional diffeomorphisms. Indeed, the definition of the
regularized boundary $X^\mu(\sigma)=(x^i=\sigma^i
,\,\,\,\rho=\varepsilon)$ will change after diffeomorphism to
$X'^\mu(\sigma )=(x'^i=x'^i(\sigma,\varepsilon),
\rho'=\rho'(\sigma,\varepsilon))$ where
$x'^i(x,\rho),\rho'(x,\rho)$ are diffeomorphism functions inverse
to those defined in \r{33}. But the boundary term depends on the
$d+1$ dimensional covariant divergence of the normal vector
$D'_{\mu}n'^{\mu}=D_{\mu}n^{\mu}$ and on the induced metric
$\gamma_{ij}(\sigma)$ which is also invariant:
\begin{equation}\label{40}
\gamma_{ij}(\sigma)=G_{\mu\nu}\frac{\partial X^\mu}{\partial
\sigma^i}\frac{\partial X^\nu}{\partial
\sigma^i}=G'_{\mu\nu}\frac{\partial X'^\mu}{\partial
\sigma^i}\frac{\partial X'^\nu}{\partial \sigma^i} .
\end{equation}
Thus we can deduce that there is no contribution to the cocycle
from the boundary term, just as there is no contribution from this
term in the anomaly. Then we can concentrate on the pure bulk
contribution in the effective action:
\begin{equation}\label{41}
W^{bulk}_{reg} =\frac 1{2k_{d+1}^2}\int d^{d}x\int_{\rho \geq
\varepsilon}d\rho\frac{d}{\rho^{{d\over 2} +1}} \sqrt{g(x,\rho )} .
\end{equation}
After integration in $d=2,4$ we obtain the divergent part of the
action:

\begin{eqnarray}
  W^{bulk}_{d=2}&=&\frac 1{2k_{d+1}^2}\int d^{d}x
  \sqrt{g_{(0)}(x)}\left(\frac{2}{\varepsilon}-
  \ln{\varepsilon}\frac{1}{2}R\right) ,\label{42}\\
W^{bulk}_{d=4}&=&\frac 1{2k_{d+1}^2}\int d^{d}x
  \sqrt{g_{(0)}(x)}\left(\frac{2}{\varepsilon^2}+
  \frac{1}{3\varepsilon}R-\ln{\varepsilon}
  \left(\frac{1}{24}R^2-\frac{1}{8}R^{ij}R_{ij}\right)\right) .\label{43}
\end{eqnarray}
We can now apply  diffeomorphisms \r{33} to the bulk effective
action \r{41} and calculate the cocycle of the Weyl group because
this transformation reproduces on the boundary the usual Weyl
transformation of the boundary metric. For this one should note
that the integrand of \r{41} is the covariant volume $\sqrt{G}$,
so that we get the following transformation of our
parameters.Instead of initial coordinates $x, \rho $ , metric
$g(x, \rho)=g[g_{0}(x, \rho)]$ and integration limit
$\rho=\varepsilon$, we have now $x', \rho',g'(x',
\rho')=g'[g'_{0}(x', \rho')=e^{2\sigma(x')}g_{0}(x')]$ and an $x'$-dependent
integration limit $\rho'=f(x',\varepsilon)$, where $f(x',\varepsilon)$ is the
solution with respect to $\rho'$ of the equation
\begin{equation}\label{44}
  \varepsilon=\rho(x',
  \rho')=\rho' e^{-2 \sigma(x')} + \sum_{k=2} a_{(k)}(x') \rho'^k
\end{equation}
which follows from \r{33} to \r{37}.
 The solution of \r{44} is
\begin{eqnarray}
\rho'&=&f(x',\varepsilon)=\varepsilon e^{2 \sigma(x')} +
\sum_{k=2}b_{(k)}(x')\varepsilon ^k ,\label{45}\\
b_{(2)}(x')&=&-a_{(2)}(x')e^{6\sigma(x')}=\frac{e^{2\sigma(x')}}{2}\partial_{i}
\sigma(x')\partial_{j}\sigma(x')g^{ij}_{(0)} ,\label{46}\\
b_{(3)}(x')&=&2a^2_{(2)}(x')e^{10\sigma(x')}-a_{(2)}(x')e^{8\sigma(x')}\nonumber\\&=&
\frac{e^{2\sigma(x')}}{4}\left( \frac{5}{4}(\partial\sigma )^4 -
\partial_{i}
\sigma(x')\partial_{j}\sigma(x')g^{ij}_{(2)}\right) ,\label{47}\\
b_{(4)}(x')&=&\cdots  .\nonumber
\end{eqnarray}
Finally we have to calculate the cocycle by performing the following
transformation of the divergent part of the bulk effective action
\begin{eqnarray}\label{48}
  S(\sigma,g_{(0)})&=&\lim_{\varepsilon  \to 0}
  \left[\frac 1{2k_{d+1}^2}\int d^{d}x'\int_{\rho' \geq
f(x',\varepsilon)}d\rho'\frac{d}{\rho'^{{d\over 2} +1}}
\sqrt{g'(x',\rho' )}\right.\nonumber\\&-&\left.\frac
1{2k_{d+1}^2}\int d^{d}x\int_{\rho \geq
\varepsilon}d\rho\frac{d}{\rho^{{d\over 2} +1}} \sqrt{g(x,\rho
)}\right] .
\end{eqnarray}
Actually for calculation of the transformed action in $d=2,4$ we
only have to replace in \r{42} to \r{43} $\varepsilon$ by
$f(x',\varepsilon)$ and $g_{(0)}$ by $g'_{(0)}$. We use the
following formulas
\begin{eqnarray}
\frac{1}{f(x',\varepsilon)}&=&e^{-2\sigma(x')}\left[
\frac{1}{\varepsilon}-\frac{1}{2}\partial_{i}
\sigma(x')\partial_{j}\sigma(x')g^{ij}_{(0)} +{\cal O}(\varepsilon)\right],\label{49}\\
\frac{1}{f(x',\varepsilon)^2}&=&e^{-4\sigma(x')}\left[\frac{1}{\varepsilon^2}
-\frac{1}{\varepsilon}\partial_{i}
\sigma(x')\partial_{j}\sigma(x')g^{ij}_{(0)}\right.\nonumber\\
&+&\left.\frac{1}{2}\left(\frac{1}{4}(\partial \sigma )^4
+\partial_{i}
\sigma(x')\partial_{j}\sigma(x')g^{ij}_{(2)}\right)+{\cal
O}(\varepsilon)\right],\label{50}\\
{\bf a}'_{(2)}(g'_{(0)})&=&e^{-2\sigma}\left({\bf
a}_{(2)}(g_{(0)})+\Box\sigma\right),
\,\,\, d=2,\label{51}\\
{\bf a}'_{(4)}(g'_{(0)})&=&e^{-4\sigma}\left[{\bf
a}_{(4)}(g_{(0)})- \frac{1}{2}\nabla_{i}\left(\left(
R_{(0)}^{ij}-\frac{1}{2}R_{(0)}g^{ij}_{(0)}\right)\partial_{j}\sigma
\right.\right.\nonumber\\&+&\left.\left.\frac{1}{2}\nabla^j(\partial\sigma)^2
-\nabla^i\sigma\Box\sigma-\nabla^i\sigma(\partial\sigma)^2\right)\right]
\quad d=4.\label{52}
\end{eqnarray}
We find that in \r{48} all divergences will cancel for both
$d=2,4$ cases (which is a nice indication of the correctness of
our idea) and we obtain the following finite expressions for
\r{48} in $d=2,4$
\begin{eqnarray}
S^{d=2}(\sigma,g_{(0)})&=&\frac 1{2k_{3}^2}\int
d^{2}x\sqrt{g}(g^{ij}\partial_i \sigma
\partial_j \sigma - R_{(0)} \sigma) ,\label{53}\\
S^{d=4}(\sigma,g_{(0)})&=&\frac 1{2k_{5}^2}\int
d^{4}x\sqrt{g}\left[\frac{1}{4}(\partial \sigma)^4
+\frac{1}{2}\Box\sigma (\partial\sigma )^2
-\frac{1}{2}\left(R_{(0)}^{ij}-\frac{1}{2}R_{(0)}g_{(0)}^{ij}\right)\partial_i\sigma
\partial_j\sigma\right.\nonumber\\
&-&\left.\frac{1}{4}\left(\frac{1}{3}R_{(0)}^2 -
R_{(0)}^{ij}R^{(0)}_{ij}\right)\sigma\right] .\label{54}
\end{eqnarray}
These expressions are in full agreement with \r{17} and \r{23} after using the
standard relation $\frac 1{2k_{3}^2}=\frac{c}{24\pi}$ and the $5d$
gravitational constant \r{32}. Thus we have shown that finite
diffeomorphisms \r{33} applied to bulk contributions of the AdS
effective action generate cocycles of the Weyl group in dimensions
$2$ and $4$ with the right form.

\section{Conclusion}
Above we considered the 1-cocycle of the Weyl group (i.e. integral
trace anomaly) in the AdS/CFT approach. We have shown that special
$d+1$ dimensional diffeomorphisms \cite{Theis} of the finite form
\cite{S} applied  to the divergent part of the AdS action
originating from the bulk integral reproduce the 1-cocycle of the
Weyl group corresponding to the correct anomaly in $d=2,4$. It
will be interesting to extend this consideration to $d=6$. This
will be interesting because the renormalization properties of
anomaly coefficients in $AdS_7/CFT_6$ still lack explanation
because we still do not have a good explanation of the
renormalization of the Euler density coefficient from the weak to
the strong-coupling regimes \cite{BFT} for the (2,0) tensor
multiplet, like the difference in the $R$-symmetry anomaly
structure for this multiplet in weak and strong-coupling regimes
\cite{Min}. Thus any investigation in the field of anomalous
behavior in the AdS/CFT picture will be interesting and important.
\section{Acknowledgements}
R. Manvelyan acknowledges financial support of DFG (Deutsche
Forschungsgemeinschaft). This work was also partially supported by
INTAS grant No:99-0590.

\end{document}